\documentclass[alpha-refs]{wiley-article}

\usepackage{siunitx}
\usepackage{xcolor}
\usepackage{textcomp}
\usepackage{blkarray}
\usepackage{soul}

\robustify{\ref}

\papertype{Original Article}
\paperfield{Journal Section}

\title{Beyond skill scores: exploring sub-seasonal forecast value through a case study of French month-ahead energy prediction.}

\author[1]{Joshua Dorrington}
\author[2]{Isla Finney}
\author[1]{Tim Palmer}
\author[1,3]{Antje Weisheimer}

\affil[1]{Atmospheric, Oceanic and Planetary Physics, University of Oxford, Oxford, United Kingdom}
\affil[2]{Lake Street Consulting Ltd}
\affil[3]{European Centre for Medium-Range Weather Forecasts, Reading, United Kingdom}
\corraddress{Josh Dorrington, Atmospheric, Oceanic and Planetary Physics, Clarendon Laboratory, University of Oxford, Sherrington Rd, Oxford OX1 3PU}
\corremail{joshua.dorrington@physics.ox.ac.uk}

\fundinginfo{Josh Dorrington is funded by NERC Grant NE/
L002612/1 }

\runningauthor{J. Dorrington et al.}

\begin{document}

\maketitle

\begin{abstract}
We quantify the value of sub-seasonal forecasts for a real-world prediction problem: the forecasting of French month-ahead energy demand. Using surface temperature as a predictor, we construct a trading strategy and assess the financial value of using meteorological forecasts, based on actual energy demand and price data. We show that forecasts with lead times greater than 2 weeks can have value for this application, both on their own and in conjunction with shorter range forecasts, especially during boreal winter. 

We consider a cost/loss framework based on this example, and show that while it captures the performance of the short range forecasts well, it misses the marginal value present in medium range forecasts. We also contrast our assessment of forecast value to that given by traditional skill scores, which we show could be misleading if used in isolation. We emphasise the importance of basing assessment of forecast skill on variables actually used by end-users.

\keywords{sub-seasonal prediction, forecasting, applications, energy}
\end{abstract}

\section{Introduction}\label{sec:intro}
Over the last 15 years operational forecasting centres are increasingly extending forecasts into the 3-6 week timescale, often called the monthly, extended or sub-seasonal range. Driven by the notion of seamless prediction [\cite{Hoskins2013}], the aim is to fill the gap between conventional 2-week weather forecasts and longer term seasonal projections. This can be done by harnessing the  variability of slow drivers such as sea ice [\cite{Chevallier2019}], the land surface [\cite{Dirmeyer2019}], and atmospheric-oceanic processes such as the Madden-Julian Oscillation [\cite{Vitart2017}].

Despite sub-seasonal forecasting becoming well established, the applications and interpretation associated with these extended forecasts can still be unclear, both because robust skill is often only achievable for long time-averages of variables, and because the nature of forecast skill becomes inherently more probabilistic at longer lead times. End users will get more value from forecasts if we can clearly understand the variables and timescales that sub-seasonal forecasts can predict well, and where they have more limited applicability (see \cite{White2017} for a detailed discussion). However, much of the literature focused on verifying sub-seasonal forecast skill uses either mid-tropospheric, large scale fields [\cite{Buizza2005}], or spatially localised station data [\cite{Monhart2018}], both of which are somewhat removed from end-user application which in many cases is interested in national averages at high temporal resolution.

Further to this, it is not always clear how directly forecast skill as measured by the usual metrics maps to actual, realisable value to an end user. This distinction underlines a gap that often exists between those who develop and produce forecasts and those who ultimately use them to make decisions. The importance of linking forecast development to user needs and of addressing the incommensurability of perspectives between these two groups, one of which speaks of correlation skill, and another annual expected return, was emphasised in \cite{Palmer2002}. Their figure 1 captures schematically the idea of linking forecast verification metrics directly to end-user benefits. More recently, an extensive report on the future development of sub-seasonal forecasts [\cite{NSC2016}] made the the same point: their key findings 3.8 and 3.10 highlight the importance of developing dialogue and explanations of reasoning between model developers and users, while their recommendation B explicitly calls for stakeholder involvement in forming verification metrics. In short, we must recognise our models are only as good as end-users find them to be.

In this work we focus on this question of skill scores, taking a specific case study of French energy markets as an example, and we compare progressively more idealised metrics of value to highlight the different conclusions they might imply if used in isolation. Precisely, we examine the problem of French month-ahead energy forecasting over a 9-year period. The need for improved analysis of subseasonal forecasts for the energy sector was emphasised in \cite{Orlov2020} who remark that 'the economic value of S2S forecasts has not been sufficiently investigated', making our choice of case study particularly relevant.

There has recently developed a valuable body of literature addressing the use of subseasonal and seasonal forecasts, many specifically targeted for energy sector applications. \cite{Soret2019} demonstrated that for some extreme event case studies, use of ensemble forecasts for surface temperature and wind speed in the mid-latitudes had potential to improve decision making even at week 4, while \cite{VanDerWiel2019} looked at subseasonal skill through the lens of weather regimes, showing they are associated with different distributions of energy-relevant variables such as demand and renewable generation. Explicit considerations of economic forecast value are found in \cite{Dutton2013} and \cite{Emanuel2012} for idealised hypothetical businesses based on trading on seasonal temperature terciles, and hurricane activity respectively. \cite{Lynch2014} goes further and presents a detailed analysis of wind energy predictability that shows skill in average wind speed over days 14-21, and possible increased returns of 60\% from the utilisation of forecasts in decision making for a realistic power-generation business case.

We aim to build on this basis, with the distinction that we look at the use of subseasonal forecasts at daily resolution, which is more desirable than weekly means for some users, and has not been often looked at beyond two week lead times in this context. A main difference from prior work in this area is that while we offer insights into the potential value of long range forecasts for the French energy market, our primary focus is to emphasise the value of assessing model skill in a user-centred context.

 Section \ref{sec:data} discusses the details of the forecast systems and data pre-processing used in this study. In section \ref{sec:value} we explain the use case of energy stakeholders trading power on the French energy market. We directly compute the value of current sub-seasonal forecasts in this case, using real-world power price and demand data. In section \ref{sec:PEV} we extract the users' cost-loss ratio from the energy example, and so can compute the commonly used metric of potential economic value (PEV). This allows us to move backwards to a more abstracted skill score, and to then directly compare the PEV to our more realistic value framework, in order to assess the validity of the underlying assumptions. Moving to purely meteorological scores typically used by the forecast verification community in section \ref{sec:skill_scores}, we examine several common skill metrics for daily temperature forecasts. We can then see how our perspective on forecast value is altered depending on the verification method used. We finish in section \ref{sec:discussion} by commenting on the implications of these results, both for using sub-seasonal forecasts for energy, and for the way in which academic meteorology assesses forecast skill.

\section{Data}\label{sec:data}

We analyse hindcast data from three different operational forecasting systems, covering the common period 1999-2018. In section \ref{sec:value}, we only use forecasts for which corresponding energy data is also available, restricting us to 2010-2018. SI figures 1-4 assess whether any significant difference in skill can be seen between 1999-2010 and 2010-2018, and we observe only a slight increase in week 1 probabilistic skill. Bearing this minor difference in mind, we consider results calculated from the full and restricted datasets comparable to each other. The  extended range forecasting system (hereafter EC45) is initialised twice weekly as a seamless continuation of the ECMWF 15-day forecast, using the same model cycle and coupled to the NEMO ocean model, running for 46 days. We use the model cycle IFS-45R1, one cycle behind the current operational cycle 46R1 at time of writing, in order to  have a full hindcast dataset available. The first 15 days of the EC45 system are run at a spectral resolution of TCo639, and upscaled after that to TCo319, corresponding to ~18km and 36km respectively. Data from the EC45 system for the 2018 boreal Summer was not available, representing the only departure from the common data period.

The ECMWF seasonal forecasting system, SEAS5, is initialised once per month using an older model cycle (CY43R1), but runs out to 7 months, at the same TCo319 resolution as the extended range EC45 forecast [\cite{Stockdale2018}; \cite{Johnson2019}]. We make use of the first 45 days of the SEAS5 data as a baseline against which to evaluate the utility of the EC45 system's increased initialisation frequency and model improvements.

To capture differences in sub-seasonal forecasts between modelling centres we also include the Environmental Modelling Centre's sub-seasonal ensemble forecast system (hereafter SubX), which is being run weekly in real time out to 35 days on an experimental basis as part of the SubX project [\cite{subx}], with a comparable spectral resolution of T574 (~33km). This project collates a number of different sub-seasonal hindcasts from different centres, however we chose not to use other model contributions due to limited hindcast periods and/or small ensemble sizes.

Analysis is performed on both annual and boreal winter (Dec through Feb, or DJF) datasets. The number of initialisation dates for each season, and ensemble size are summarised in table 1. To validate the forecasts we make use of the ERA5 reanalysis [\cite{era5}]. 

In all cases, we analyse daily 12:00 2-metre temperature data, on a 1 degree grid  averaged over a box centred on mainland France. A gridpointwise, monthly and lead time dependent bias correction was applied, by mapping quantiles of each forecast's temperature distribution to those of ERA5 during the same time period. A ’drop one out’ approach was used: for each hindcast year, data from only the other 19 years were used to bias correct. This prevents over-fitting, as long as we accept the mild assumption that the year-to-year correlation of surface temperatures is negligible. A gridpointwise sinusoidal fit to the annual cycle was computed for ERA5 using the first 4 harmonics and then removed from all datasets as was a weak linear trend, to leave stationary anomaly fields.The area averaging was then performed by applying a land-sea mask and taking a cosine-latitude weighted mean over the region [5W-8E,42N-51N], to produce the final scalar anomaly field. This region includes a small number of gridpoints from outside the borders of France, but comparisons using the ERA5 data show negligible differences in the T2m series obtained with a more detailed mask (see SI figures 5-7). 

\section{Forecast value in an energy framework}\label{sec:value}

\subsection{Methodology}\label{subsec:method}
In France as well as many European countries, energy to be delivered on a future date can be bought and sold through a liberalised national market. Power providers and traders are of course interested in buying these energy futures for an optimal price, a quantity which has many complex drivers such as macroeconomic factors, the price of fossil fuels which determines which power plants are operational, changes in installed capacity of renewables, and of course meteorological conditions (see \cite{UKERC2019}) for a discussion of modelling non-meteorological factors on the inter-annual scale). We focus on baseline energy contracts, where energy is to be delivered throughout the day of interest, avoiding the issue of high-frequency sub-daily variability in the energy demand. 

\begin{figure}
    \centering
    \includegraphics[width=\linewidth]{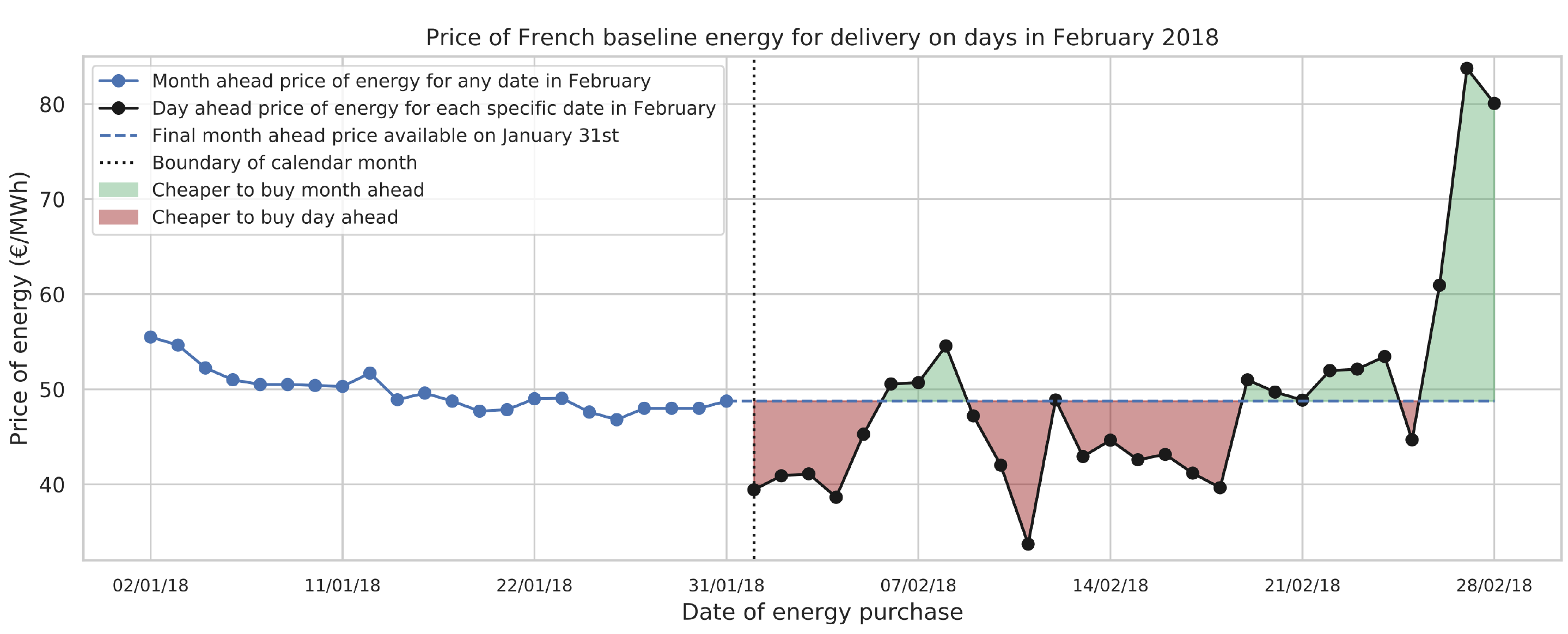}
    \caption{An example using real data of the evolution of energy prices for some given future date, in this case days in Feb 2018. During early January, the month-ahead February price (which is the same for all days in February) is the only power contract traded (solid blue line) in enough size to give a reliable benchmark of electricity cost. This varies day-to-day for a number of reasons including weather forecast changes, changes in scheduled availability of power plants, and varying coal and gas prices. As we go into February, days are priced and traded individually. A good simplification which sees real world use is to consider only the day-ahead price, (black line) once the month has begun. This has increased volatility compared to the month ahead pricing, reflecting the day to day variations in both power demand and supply. If the price spikes high, then buying at the end of the previous month is cheaper (shaded green), while if the price drops low, then buying at day ahead price is cheaper (shaded red).}
    \label{fig:price_schematic}

\end{figure}

Energy contracts tend to be traded on discrete timescales (quarterly, monthly, daily etc.). Taking the example of a month-ahead contract, the price of energy on a given day will be constant for the whole month of interest. For example in late January, a MWh of electricity purchased for February 1st is priced the same as a MWh for February 28th. Electricity within the current calendar month however is priced on a daily basis, leading to an increased sensitivity to meteorological conditions. While both these prices will evolve over time and a number of different timescales exist, we consider only two prices of interest as a first approximation; the final month-ahead price, and the day-ahead price, as depicted schematically in figure \ref{fig:price_schematic}. The question then is should energy be bought at the stable, close-to-climatology month ahead price, or at the more volatile day ahead price?

We would like to develop a plausible real-world assessment of the value of sub-seasonal surface temperature forecasts, while keeping the meteorological element front and centre. To do this we make use of real-world French energy price data covering the period 2010-2018, both month-ahead and day-ahead spot prices (calculated at around 1100 GMT the day before the target date) [\cite{french_day_ahead}]. We also use daily averaged French demand data covering the same period [\cite{french_demand}]. The economic value we obtain here will be by necessity a lower-bound, as we are ignoring the economic and societal extreme events can have on the energy sector beyond price fluctuations. As one example, extended cold spells in DJF 2016/2017 led to a series of French nuclear plants having to come offline for emergency maintenance [\cite{ENTSOG2017}].

Modelling the evolution of energy prices is a discipline in and of itself, and we are not focused on producing a high-fidelity price model in this work. Instead we build a 'good enough' trading strategy that will allow us to provide at least a lower bound on the usefulness of forecast data in this sector. Extensive prior work shows that surface temperature is the dominant driver of energy demand, with second order factors including cloud cover, rainfall, wind speed, and especially in warmer climates relative humidity and wet bulb temperature [\cite{Thornton2016}; \cite{Cho2013a}; \cite{Psiloglou2009}; \cite{Lam2008}]. We focus only on temperature, and use anomalies of ERA5 daily surface temperature with respect to monthly means to derive a weakly quadratic relationship between temperature and demand anomalies for each season. We find that despite a whole host of missing factors, ERA5 T2m makes a reasonable single-variable predictor of demand with an annual correlation of 0.66, especially in winter when the demand for heating is highest, and so consequently the relationship between temperature and demand is strongest. Appendix A shows details of the fitted relationships, and the resulting demand estimates for each season.

We start from the basis that the anomalous demand calculated with respect to month is strongly correlated with the day-ahead price but not with the month-ahead price (not shown). This means that if we can predict the energy demand from forecasts of surface temperature, then we can also predict whether it is cheaper to buy a unit of energy at the month-ahead price, or to wait until the day before and buy at the cost for immediate delivery, known as the spot price.

Given our estimate of demand, a basic trading strategy is:
\begin{itemize}
    \item Predict the anomalous demand for a future date.
    \item If the demand anomaly is greater than a threshold $d$, buy energy at the month-ahead price.
    \item Else, buy energy the day before at the spot price.
\end{itemize}

We make a forecast of demand for every date in the record, making a decision on whether to buy at the month ahead price at the end of the previous calendar month. This means that forecasts for early in the month will be more skillful than those for the end of the month; the lead time is not fixed.
There are two simple situations we might imagine a user to be in: one where they wish to buy a set amount of energy (as for a trader for example), and one in which they wish to purchase a certain fraction of the total energy demand (as might be more relevant for an energy provider trying to prevent a production shortfall). In the first case we can can evaluate the average cost per unit of power, $C$, using this strategy over $T$ days with a predictor of anomalous demand, $P_D$, as:
\begin{equation}\label{eq:set_amount}
    C_{\text{set amount}}=\frac{1}{T}\sum^T_t \left(H[P_D(t)-d]\cdot p_{\text{month}} + H[d-P_D(t)]\cdot p_{\text{day}}\right)
\end{equation}

where $p_{\text{month}}$ is the month-ahead price, $ p_{\text{day}}$, the day-ahead price, and $H$ the Heaviside step function. In the second case, we must account for the fact that we are buying more power for likely high-demand days than for low-demand days, so we weight the cost of power by the ratio of the daily demand to the mean demand: $\Tilde{D}_t:=\frac{D_t}{\langle D\rangle}$. This gives us the average cost of buying a set fraction of power:

\begin{equation}\label{eq:set_fraction}
    C_{\text{set fraction}}=\frac{1}{T}\sum^T_t \Tilde{D}_t\cdot\left(H[P_D(t)-d]\cdot p_{\text{month}} + H[d-P_D(t)]\cdot p_{\text{day}}\right)
\end{equation}

As references we consider 4 predictors that make no use of forecast data. We take 2 'perfect' predictors, one in which we know the future demand anomaly precisely, and one in which we know only the future surface temperature perfectly, and must estimate the demand using our simple demand curve. This allows us to separate out any deficiency in meteorological skill from the drawbacks of our simplistic demand model. As climatological references, we consider the strategies of always buying at the day-ahead spot price, always buying month ahead, and a random strategy where the probability of buying day-ahead and month-ahead are both 0.5 for any given day. Buying for the day-ahead is on average cheaper than buying a month-ahead; a premium is paid for the lower volatility of the long-term contracts.

For each date in the record, we produce a demand forecast from each system using the most recently initialised forecast for which it was still possible to buy power at the month ahead price. For example a model may have three forecasts for the 9th February, initialised on the 2nd Feb, the 26th Jan and the 19th Jan, at lead times of 7, 14 and 21 days respectively. In that case while the 7-day forecast is most recent, it was initialised after the last opportunity to purchase at the month ahead price (31st Jan) and so we would use the day 14 forecast instead. If no forecast had been available we would instead use the climatological strategy of always buying on the day before. We use the forecasts' ensemble mean temperature to predict demand. Taking the ensemble mean after predicting demand made no qualitative difference.

In order to better separate out the relative value of the short and long range forecasts, we also consider cases where only the first 15 days of the forecast are available, and the case where only long range forecasts (lead times greater than 15 days) are used.  We don't consider these additional cases for the seasonal SEAS5 system, as forecasts are initialised at the beginning of every month and so only lead times $>$28 days are ever available for decision making. For the subseasonal systems the median lead times using all forecasts, only the first two weeks and only days 15+ are 18 days, 9 days and 24 days respectively. Even though this is the same for both EC45 and SubX, the SubX lead times are skewed, with many fewer forecasts for days 1 to 6 contributing due to its lower initialisation frequency.

The obvious action threshold to choose is $d=0$; we should buy on any positive demand anomaly. However due to the higher average price of month-ahead contracts, and the imperfect skill of forecasts, we also choose to consider a more cautious strategy of only buying in advance if we predict an annual upper tercile demand anomaly ($\geq54.4$ GWh).

\subsection{Results}\label{subsec:results}

\begin{figure}
    \centering
    \includegraphics[width=\linewidth]{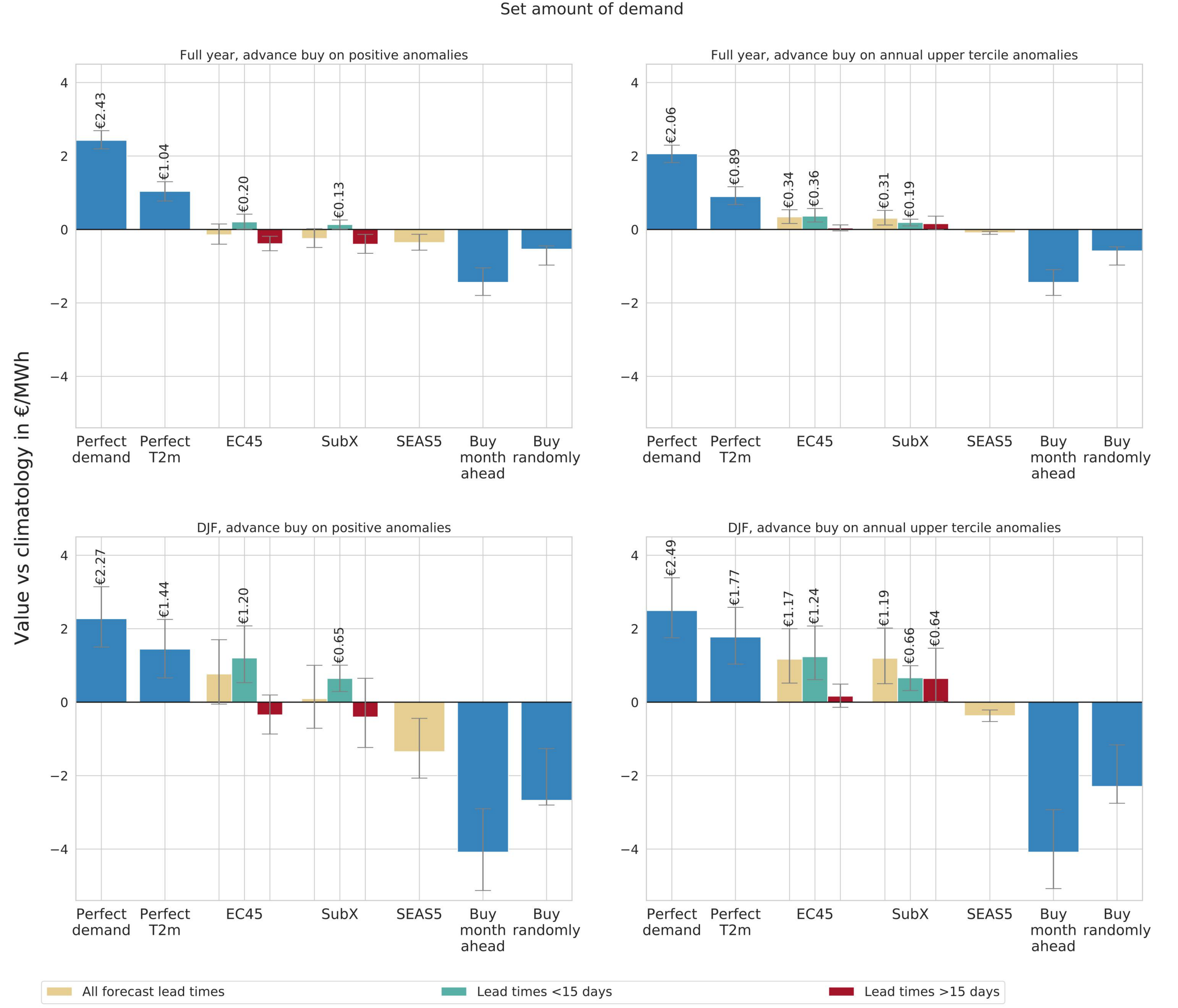}
    \caption{The average saving on a MWh of French baseline energy over the period 2010-2018 under different purchasing strategies compared to buying at the day-ahead price. Subplots \textbf{a)} and \textbf{b)} are the average cost over the full year, while \textbf{c)} and \textbf{d)} are for DJF only. \textbf{a)} and \textbf{c)} are based on buying at month-ahead price when the demand anomaly relative to the month is predicted to be positive. \textbf{b)} and \textbf{d)} are based on buying at month-ahead rate when demand anomaly is in the upper tercile. Error bars show the 5th to 95th percentiles estimated by boot-strapping over target dates. Where the 5th percentile exceeds 0, the mean saving provided from using that strategy is marked above the bar.}
    \label{fig:value_amount}
\end{figure}

In this section we present the difference in average energy price between the trading strategies discussed in section \ref{subsec:method} and the strategy of always buying energy the day before, which we note again as the best climatological strategy. In figure \ref{fig:value_amount} we look at the set energy amount scenario of equation \ref{eq:set_amount} while \ref{fig:value_amount} shows the set energy fraction case of equation \ref{eq:set_fraction}. Errors on these differences are estimated using 1000 bootstrap resamples over forecast target dates, and those strategies whose 5th percentile is positive are annotated with their associated potential saving per unit of energy. When references to statistically significant differences between forecast systems are noted, it is a result of analysis of this kind, but we don't show differences between all possible combinations of trading strategies for conciseness.

We start by looking at the value associated with a hypothetical perfect forecast of demand. Under our trading paradigm proposed here, there is the potential to save ~\texteuro2.50/MWh for a user buying a set amount of baseline energy, while for a user buying a constant fraction of daily energy generation this rises to ~\texteuro2.80/MWh on an annual basis and up to \texteuro3.86/MWh during winter. With the average cost of power on the order of \texteuro45/MWh, these are hypothetical price savings of roughly 5\%, which would certainly be of interest to industry stakeholders. For the DJF increased fraction case, the price is much higher due to the increased effect of outliers, ~\texteuro62/MWh, and so also has a maximum saving of roughly 5\%.

Of course our demand model can not meet these theoretical targets, even when using observational ERA5 temperatures as input. Perfect forecasts for DJF come closer to the theoretical maximum saving than perfect annual forecasts, realising 71-75\% of the maximum saving depending on price scenario, and 43-54\% respectively. This is due simply to the much better temperature-demand fit obtained for DJF than other seasons, while the improved value obtainable for the set fraction case is due to the increased influence of high demand days under this scenario; as our model captures only some of the variability, we are more able to predict the sign of strong outlier events than weak ones.

Looking now at the performance of our forecast systems, we immediately see that forecasts for the full year hold much less value than during DJF, which again follows quite naturally from the better temperature-demand fit for DJF than other seasons. However we do see statistically significant, albeit low, value on an annual basis for both forecast systems, for both the set amount and set fraction cases. The high cost penalty associated with a random trading strategy means that when looking at upper median trades, longer range forecasts are indistinguishable or robustly worse, than always buying energy the day ahead, and their presence removes value from the forecast system. When only the first 15 days of forecasts are used however, both EC45 and SubX have statistically significant savings associated with them, which while no larger than \texteuro0.50/MWh on an annual basis, are between \texteuro0.65/MWh and \texteuro2.03/MWh during DJF. For the set fraction case DJF, EC45 even shows overall skill, although the longer range forecasts are indistinguishable from climatology.

The first 15 days of EC45 are significantly better than for SubX for the set fraction case during DJF as a result of more regular forecast initialisation. We also see quite clearly that for the paradigm described here, there is no value in the SEAS5 system, despite, as we will show in later sections, it having T2m skill comparable to the subseasonal systems. This is purely due to the fact it is always initialised on the 1st of the month; the worst possible time from an energy perspective! This is made explicit in figure \ref{fig:lead_time} which shows the cumulative fraction of forecasts used in this section for increasing lead time, and where the shortest lead time SEAS5 forecast used is a 28 day forecast.

 If we look at strategies using the more conservative upper tercile threshold, we see the value of forecasts goes up both annually and for DJF. The forecasts beyond 15 days for the SubX system now have significant value on their own, in all scenarios except for the annual set amount case, and while the EC45 system does not show the same long range skill, it is not significantly worse than the SubX system. The higher value of the first 15 days of EC45 compared to SubX is significant however, for the same reason as before, it's increased initialisation frequency. Crucially however, in this case the presence of the longer range forecasts substantially improves the quality of the SubX system as a whole, making it equally skillful to EC45, and during DJF almost doubling the associated cost saving under both user cases.

In summary the simple demand model and trading strategy used here certainly limits the possible value temperature forecasts can provide in this framework, however it provides enough skill to show that under some conditions forecasts can have value for daily prediction at lead times exceeding 2 weeks. Results were optimal when trading on upper tercile anomalies, and tuning the action threshold more precisely could conceivably increase value yet further. We also 
In summary, the simplicity of the temperature-based demand model used here limits the value of the forecast information, but regardless is good enough to demonstrate that (some) forecasts have value. Results were optimal when trading on upper tercile anomalies, and tuning the action threshold more precisely could conceivably increase value yet further. Forecasts give better savings in general for DJF than for the full year, as a result of both higher surface temperature skill and a closer temperature-demand relationship. Increased initialisation frequency of course makes a forecast system more valuable as we saw for the comparison between the first two weeks of EC45 and SubX, but the addition of forecasts $\geq$ 15 days can make up for that lower initialisation frequency, at a lower computational cost. We now proceed to examine how closely these observations match up to the implications provided by more traditional, meteorological skill frameworks

\section{Potential Economic Value}\label{sec:PEV}

We have seen above that under realistic decision making conditions, longer range temperature forecasts can provide value in some cases for end users in the energy sector. We would also now like to consider how closely the value of forecasts shown above can be emulated within the more traditional cost/loss framework of the potential economic value (PEV) [\cite{Murphy1985}; \cite{Sultan2010}]. The main simplifications involved in moving to the PEV framework is that the cost and loss no longer depend on the magnitude of the demand continuously, only on whether the action threshold is exceeded or not, and that we switch from looking at the predictability of demand directly, and use T2m as a proxy.

The PEV can be defined in terms of the confusion matrix, $M^{\text{(conf)}}$, which describes how likely we are to predict or miss an impactful event $e$, and the cost matrix, $M^{\text{(cost)}}$ which tells us the associated cost of acting on our forecast:

\[M^{\text{(conf)}}=
\begin{blockarray}{ccc}
 & \text{Predicted} & \text{Not Predicted} \\
\begin{block}{c(cc)}
  \text{Event}& P\text{\tiny{True Positive}} & P\text{\tiny{False Negative}}  \\
  \text{No Event}& P\text{\tiny{False Positive}} & P\text{\tiny{True Negative}}\\
\end{block}
\end{blockarray}
,\hspace{3mm}
M^{\text{(cost)}}=
\begin{blockarray}{ccc}
 & \text{Action}&\text{No Action} \\
\begin{block}{c(cc)}
\text{Event}&C & L  \\
  \text{No Event}&C &0 \\
\end{block}
\end{blockarray}
\]

Where $P$ is probability, $C$ is the cost of taking action (i.e. buying month-ahead energy), and $L$ is the loss associated with inaction, in our case paying a hefty day-ahead price for power during a period of high demand. The 4 probabilities in $M^{\text{(conf)}}$ are estimated using our demand forecasts and observational data from the previous section. If the elements of these matrices can be defined and calculated, then the PEV is given by:

\begin{equation}
    PEV=\frac{C_{\text{clim}}-\sum_{i,j}M^{\text{(conf)}}_{ij}M^{\text{(cost)}}_{ij}}{C_{\text{clim}}-C\cdot P(e)}
\end{equation}
where
 \begin{equation}
    C_{\text{clim}}= \begin{cases}
C &\text{if } C/L < P(e)\\
L\cdot P(e) &\text{otherwise.}
\end{cases}
\end{equation}

This is simply the cost saving acting on the forecast provides, expressed as a fraction of the difference between the climatological cost, $C_{\text{clim}}$, and the cost unavoidable with even perfect knowledge, $C\cdot P(e)$ where $P(e)$ is the probability of the event (i.e. it ranges from 1 for a perfect forecast to 0 for a skill-less one. We are in a position to directly extract the ratio from our applied scenario above, and do this by taking the cost of action as the average amount by which month-ahead price exceeds day ahead price, and by taking the loss as the average amount by which day-ahead cost during a period of high demand exceeds day-ahead cost during low demand.  Depending on season or trading scenario chosen, we find the C/L ratio ranges from 0.6-0.7. We find our results are not qualitatively sensitive to the small differences within this range, and so use an intermediate value of 0.65 for our analysis here.

Our construction of PEV in this way is similar to the approach laid out in table 12.1 of \cite{Dutton2018}, although we have found no explicit use of it, or estimation of a cost-loss ratio in this manner, in the energy or meteorology literature. Indeed, while there are a range of studies that have looked at the potential economic value of sub-seasonal forecasts [\cite{Materia2020}; \cite{Zhu2002}; \cite{Emanuel2012}], or with a more complex trinary decision model [\cite{Sultan2010}], results are normally presented for a range of cost-loss ratios, with no comment made on what value is most realistic for a given application. The optimal PEV will always be found for a user with a cost-loss ratio equal to the probability of the event being predicted (i.e. in our case 0.5 or 0.33 depending on action threshold). As most work looks at action on upper tercile, quintile or decile events this skews the applicability of a forecast heavily towards low cost-loss ratios, and it is of course not \emph{a priori} given that the optimal user of a forecast actually exists. In our case our cost-loss ratio is quite high compared to the optimal value and we will see this reflected in relatively modest PEV values.

We find in figure \ref{fig:PEV} that there are only small differences between PEV for DJF and annually, and between action thresholds, with 1-day forecasts tending to realise 70-80\% of the potential value. The PEV falls in all cases below 40\% between days 7-10, and becomes statistically indistinguishable from 0 between days 10 and 14. We also see immediately that the large differences between models vanish, with the exception of a slightly lower annual PEV for SubX.

Based on these results in isolation, we would expect to see no real-world value in forecasts greater than 15 days, and an average PEV of approximately 0.4 over the first two weeks of a forecast, for both action thresholds and across seasons.
This matches broadly speaking with the annual values calculated in the previous section, especially for the ultimately optimal approach of trading on upper tercile events, which also showed 40\% of the potential saving provided by a perfect temperature forecast being realised. However, the savings seen for extended range forecasts are in direct contrast to the zero-value PEV, and results for DJF, where models were able to provide upward of 60\% the value a perfect temperature forecast would provide.

Therefore we see the PEV in this case misses a valuable component of the user end-case and causes an overly pessimistic estimate of forecast skill; more extreme demand anomalies have a larger price impact. This could be partially accounted for by extending the event/non-event dichotomy to include strong events, weak events etc., which would give a finer grained perspective but which also leads to a proliferation of free parameters dictating what kind of mitigative actions should be taken. Despite this limitation of the PEV we see a broad qualitative agreement between the two approaches, especially for the  annual average, with the clearest value in the first 2 weeks of the forecast, and PEV over that period falling in the 30-60\% range.

\section{Non-economic skill scores}\label{sec:skill_scores}

Having evaluated forecast value in economic terms, it is now useful to more purely meteorological skill scores, and we focus on some of the most frequently employed skill metrics in the literature. In figure \ref{fig:skill_scores} we show the Pearson correlation and root-mean-square error (RMSE) between the forecast ensemble mean and observations, and the continuous ranked probability score (or CRPS) (a measure of probabilistic skill, see \cite{Hersbach2000} for details) for daily French surface temperatures. As for the PEV we see relatively minor differences in skill between the different forecast systems, and most differences are confined to week 1 of the forecast.

Correlation skill remains significantly above zero out to days 22 and 27 for SubX and EC45 respectively, both during DJF and annually (fig \ref{fig:skill_scores}i)), while it isn't possible to distinguish robust skill in SEAS5 beyond day 15 in DJF or day 18 annually. These scores are more optimistic than any other metric, and from this we might well conclude that the daily forecast retains some marginal value into the extended range as we have realised in section \ref{sec:value}, although of course there is no way to quantify the term 'marginal' with a non-economic score. It is worth noting that despite correlations being low after week 2 (<0.3), they are still significant, and as there clearly is skill at these longer lead times shown in section \ref{sec:value}, they are also meaningful.

In contradiction, annual RMS Error saturates between days 11 and 15 (fig \ref{fig:skill_scores}ii)), which would then imply essentially no useful information existed in the extended range. The CRPS (fig \ref{fig:skill_scores}iii)) provides an equivalent message, saturating between days 15 and 20, and implying no added value could be gained even from a full probabilistic perspective. This can be best understood in light of the fact that our trading strategy in section \ref{sec:value} is based on threshold exceedance, reducing sensitivity to the differences between model and verification that would degrade RMS error and CRPS. Within the energy framework, as long as sign and approximate magnitude of anomalies are well predicted, value can be extracted.

\section{Discussion}\label{sec:discussion}

In section \ref{sec:value} we have quantified the real world value of sub-seasonal temperature forecasts for energy trading on a national market. Using a simple trading strategy, we show that compared to a climatological strategy, subseasonal forecast systems can provide savings of \texteuro 0.30-2.10/MWh, which represents ~40\% of the value a perfect temperature forecast could provide annually, and up to ~70\% during DJF. Due to the simplicity of our model, the largest added value is for DJF when demand is most sensitive to temperature, but we also find some significant value on an annual basis. Not all this value comes from the first two weeks of the forecast systems; we show that SubX forecasts beyond 2 weeks have value on their own, 40\% as much as a perfect forecast during DJF, and that without these extended range forecasts the total value of the system is reduced. We also saw clearly in this framework how differences in forecast initialisation frequency had a substantial impact on the usefulness of the system, with SEAS5 showing no value at all, despite being a very similar model to the EC45 system.

Having found these results, in section 4 and 5 we explore how well they are mirrored by more idealised/abstracted skill scores. In section 4 we have managed to extract a cost-loss ratio from our data, and compare the true forecast value to the estimate of a simplified PEV model. We find broad qualitative agreement with section 3, especially for shorted lead times, but the binary cost-loss assumption underlying the PEV causes an underestimation of forecast value in cases where cost correlates strongly with the magnitude of the weather event (i.e. the binary assumption of event/no event is simplistic). In our case this means PEV is indistinguishable after week 2, and the value in DJF is underestimated. Additionally if we had averaged over different cost-loss ratios, or assumed an optimal value as is commonly done, we may have arrived at quite different results, as the cost-loss ratio we calculated, 0.6-0.7, is much higher than for the optimal value. Simplifying our skill scores yet further, a comparison to conventional, non-economic scores highlights that even when RMS error and CRPS are at climatological levels, end users might still be able to extract value from forecasts, even with very low positive correlations. 

As we have emphasised, our estimations of value using our trading framework form only a lower bound. A more careful consideration of action thresholds, use of population weighted area means, and detailed analysis of very extreme events which affect energy supply as well as demand (as mentioned in the introduction) could increase the forecast value further, making the underestimation of value from traditional skill scores all the more severe.

 We acknowledge that we have considered in this work only a single end-user sector in a single country, and that even within this domain, other frameworks for making forecast-based decisions exist. We present these results then not in order to draw far-reaching conclusions about current operational forecast value, but to provide an argument and conceptual model for assessing sub-seasonal forecasts in a way that directly connects with end-user requirements. 

Extending the methodology of this study to additional national domains, and in general trying to use specific end-user examples to choose variables for forecast skill verification, could help shape model development and increase enthusiasm amongst forecast users. As discussed in the introduction there is a growing body of literature presenting user driven assessments of model value, with greater or lesser degrees of complexity, whose methodologies could potentially be used more routinely to inform model development. Of course there are many benefits to simple skill scores: they are easy to understand and calculate, and have general applicability, but much as the importance of a hierarchy of models is recognised as vital to atmospheric science, adopting a hierarchy of skill scores, including intermediate complexity options as we present here, may prove to be of equal use in converting science into societal benefit.

\section*{acknowledgments}

We acknowledge the National Environmental Research Council for funding of an industrial internship on which the origins of this work are based, ECMWF for making IFS hindcast data available by commercial license, and the agencies that support the open-source SubX system; NOAA/MAPP, ONR, NASA, and NOAA/NWS.

We thank Dave Macleod and Daniel Befort for valuable discussions that helped shape the direction and enhanced the overall understanding of this work. and the reviewers for their constructive comments which have been of great help in improving this manuscript.

\section*{Supporting information}

Supplementary material is available for this paper. Figures S1-S4 evidence the limited difference in T2m forecast skill between the periods 1999-2010 and 2010-2018, while figures S5-S7 show the low sensitivity of French area-averaged temperature to the precise area mask used.

\bibliography{French_T2m_paper}
\appendix
\section{Appendix}
In this appendix, the details of the fitted relationships between ERA5 surface temperature and demand, and between demand and energy price for the period 2010-2018 are shown in figures \ref{fig:fillin1} and \ref{fig:fillin2} respectively, and briefly discussed.

\subsection{Fitting temperature to energy demand}
In order to fit a relation between surface temperature and demand, we first need to remove a seasonal cycle from both time series. Different deseasonalising methods were trialled, and removing the calendar monthly mean from each daily data series was found to give the best fit. The first row of figure \ref{fig:fillin1} shows the relationship of these temperature anomalies against demand anomalies for each season, with a quadratic least squares fit shown by the blue line. Polynomial fits up to 5th order were trialled but gave no significant improvements. These quadratic fits have been used to produce ERA5 estimates of demand anomalies, as shown on the bottom row of figure \ref{fig:fillin1}. As we see, demand anomalies are most sensitive to temperature anomalies during DJF, with JJA showing a particularly weak relationship. The bimodality in JJA demand represents a weekday/weekend split, presumably due to increased commercial air-conditioning during the summer months. Remembering that the proportion of demand variance that the temperature anomaly explains is the correlation squared, we see why we were only able to obtain forecast value during DJF and on an annual basis; in Spring through Autumn, temperature explains less than 40\% of the demand variability.

\subsection{Relationship between energy demand and potential price saving}
The top row of Figure \ref{fig:fillin2} shows the correlation between daily French energy demand and the month-ahead price saving, which we define as the difference between the day-ahead price, and the final month-ahead price. The bottom row shows the same but using daily demand anomalies, calculated with respect to the mean demand for each calendar month. We see that deseasonalising most improves the fit during the transition months (i.e. MAM and SON) when the weather, and consumers' behavioural habits are changing most rapidly. We see that after deseasonalising, the relationship is relatively stable across seasons, with the correlation of the anomaly time series varying between 0.61 and 0.72, (or equivalently, explaining between 37 and 52\% of the variance in price differences). 

\begin{table}
\centering{
\small
\label{tab:data_table}
\caption{A summary of the operational forecasting systems used in this work. Ensemble sizes are for the hindcast datasets used in this analysis, operational forecast ensembles are considerably larger. For more details see the body text of section \ref{sec:data} and references therein.}
\begin{tabular}[width=\linewidth]{|c|c|c|c|c|c|c|c|}
\hline
Name & Originating & Forecast & Initialisation & No. of Annual/DJF & Time Range of & Ensemble\\
&Centre &Period Used & Frequency &Initialisations & Initialisation Dates& Size\\
\hline
 EC45   &ECMWF  & 46 days &  2/week &2146/734 & 03/01/1999 - 30/05/2018 & 11\\
 \hline
 SEAS5  &ECMWF &46 days & 1/month  & 219/78 & 01/01/1999 - 01/08/2018 & 25\\
 \hline
 GEFS   &EMC (SUBX)  & 35 days & 1/week & 1017/336  & 01/06/1999 - 30/05/2018& 11\\
 \hline
\end{tabular}
}
\end{table}

\begin{figure}
    \centering
    \includegraphics[width=\linewidth]{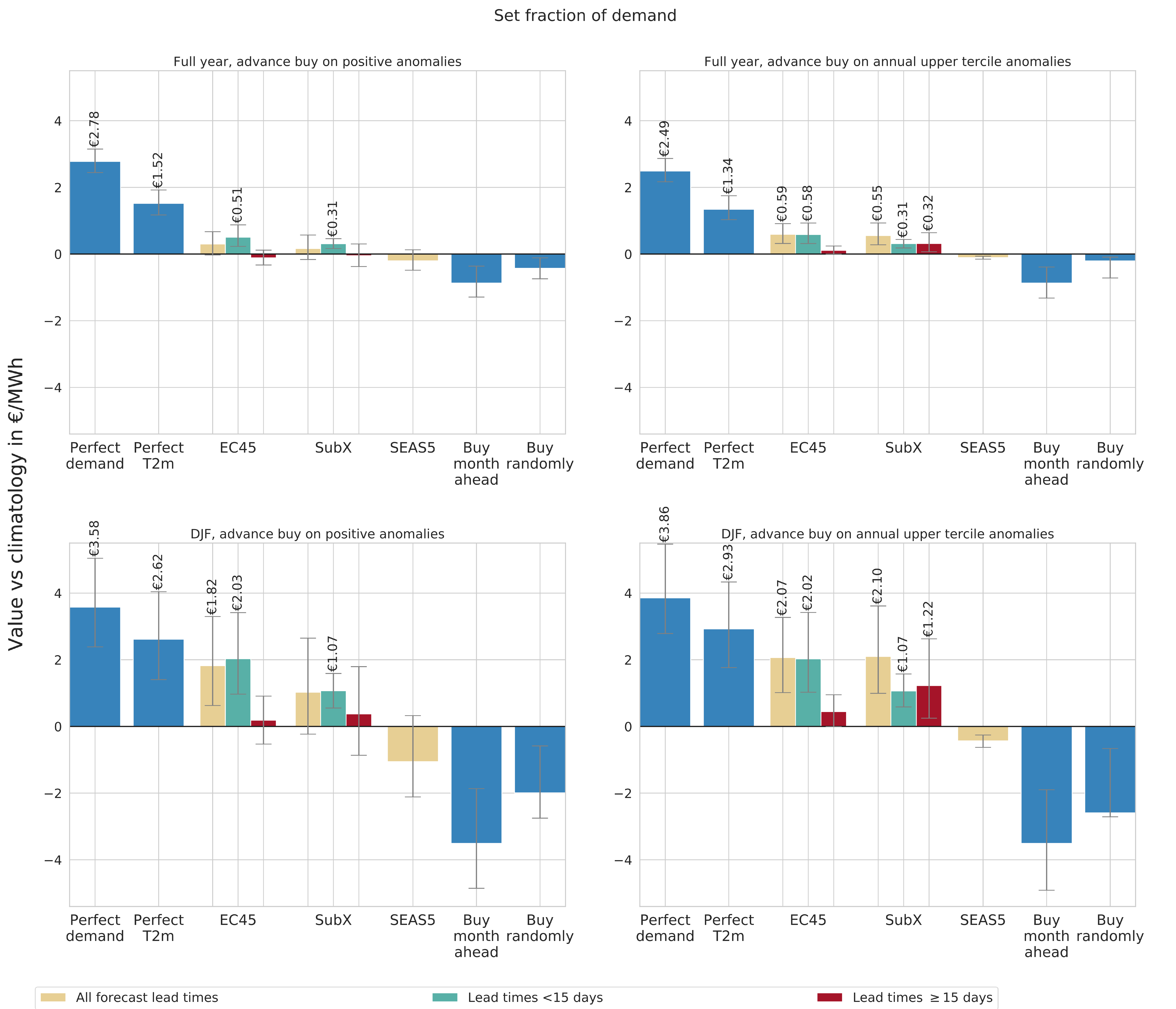}
    \caption{As for figure \ref{fig:value_amount} but showing the average saving on a MWh of French baseline energy weighted by the amount of demand on that day. This represents the scenario where a user is interested in buying a set fraction of the daily demand. Prices are explicitly valid for a user whose average energy obligation is 1 MWh.} 
    \label{fig:value_frac}
\end{figure}

\begin{figure}
    \centering
    \includegraphics[width=0.6\linewidth]{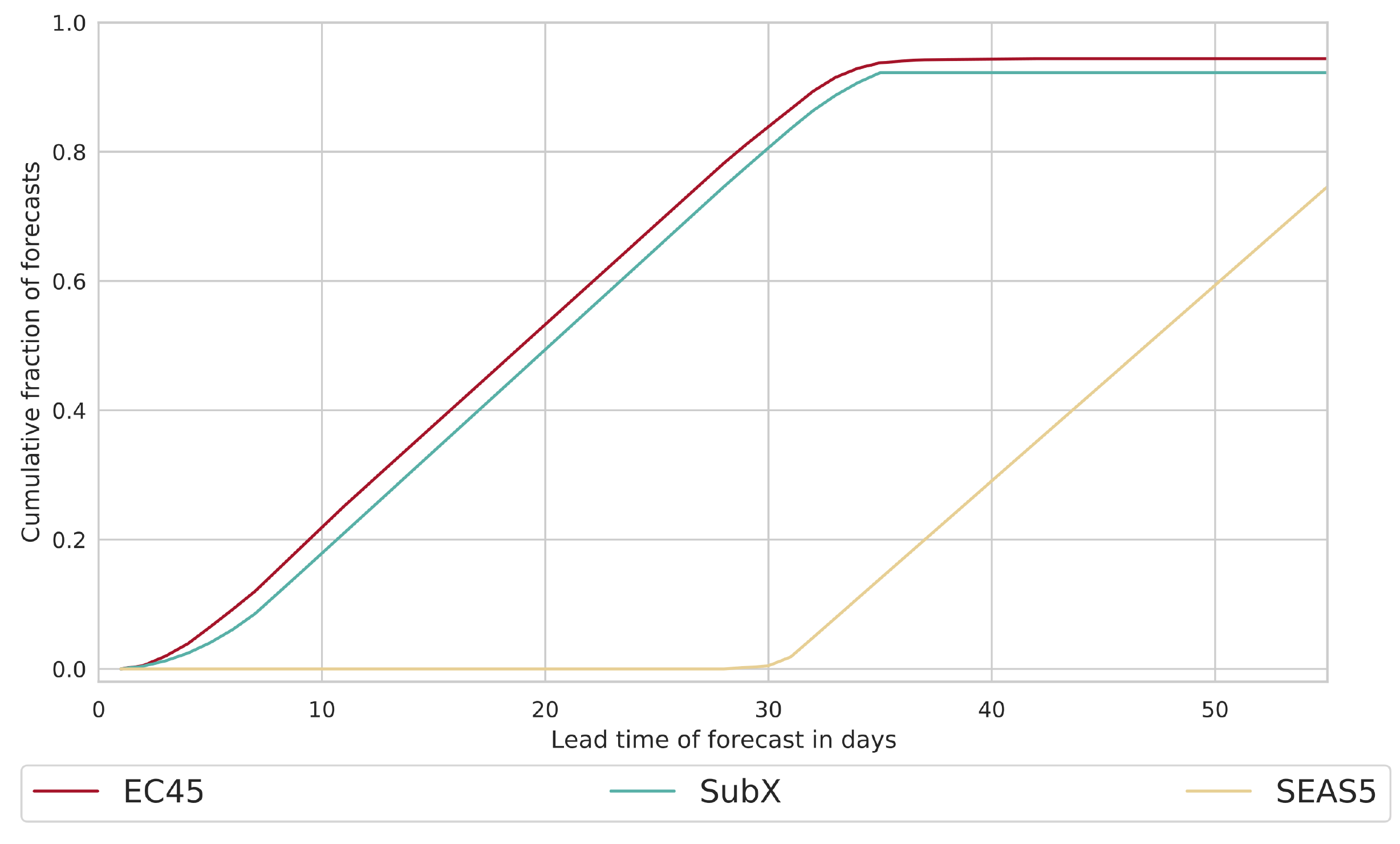}
    \caption{The cumulative distribution of forecast lead times used to make trading decisions in figures \ref{fig:value_amount} and \ref{fig:value_frac} for each forecasting system. The increased initialisation frequency of the EC45 system compared to SubX means that the average lead time of forecasts used for decision making is lower.}
    \label{fig:lead_time}
\end{figure}

\begin{figure}
    \centering
    \includegraphics[width=\linewidth]{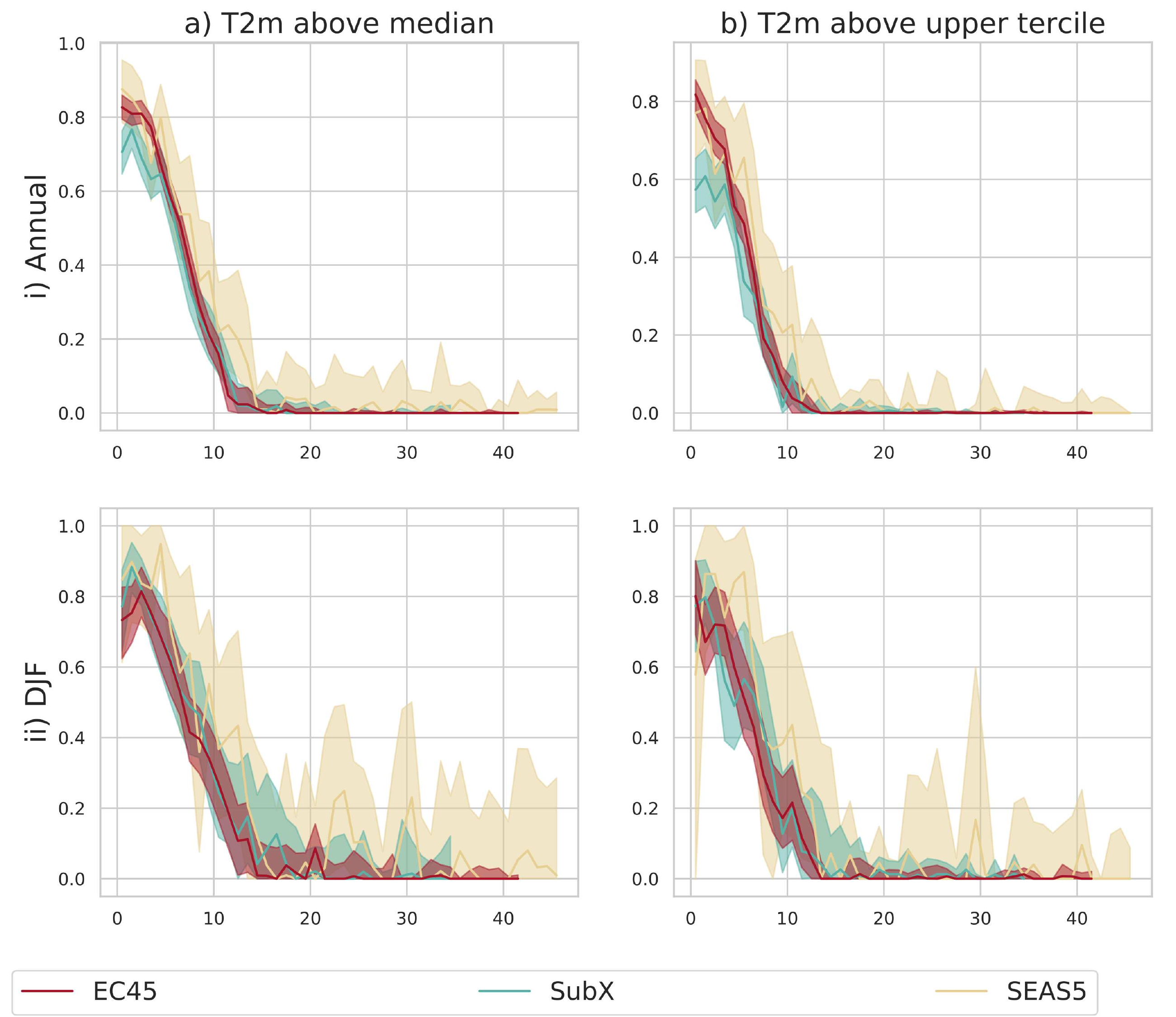}
    \caption{The potential economic value of forecasts for forecasts of \textbf{a)} demand exceeding the median, and \textbf{b)} demand exceeding the upper tercile, annually in \textbf{i)} and for DJF only in \textbf{ii)}. A cost-loss ratio of 0.65 is used in all cases. Shading shows the first standard deviation estimated by bootstrap resampling of forecast days. }
    \label{fig:PEV}
\end{figure}

\begin{figure}
    \centering
    \includegraphics[width=0.95\linewidth]{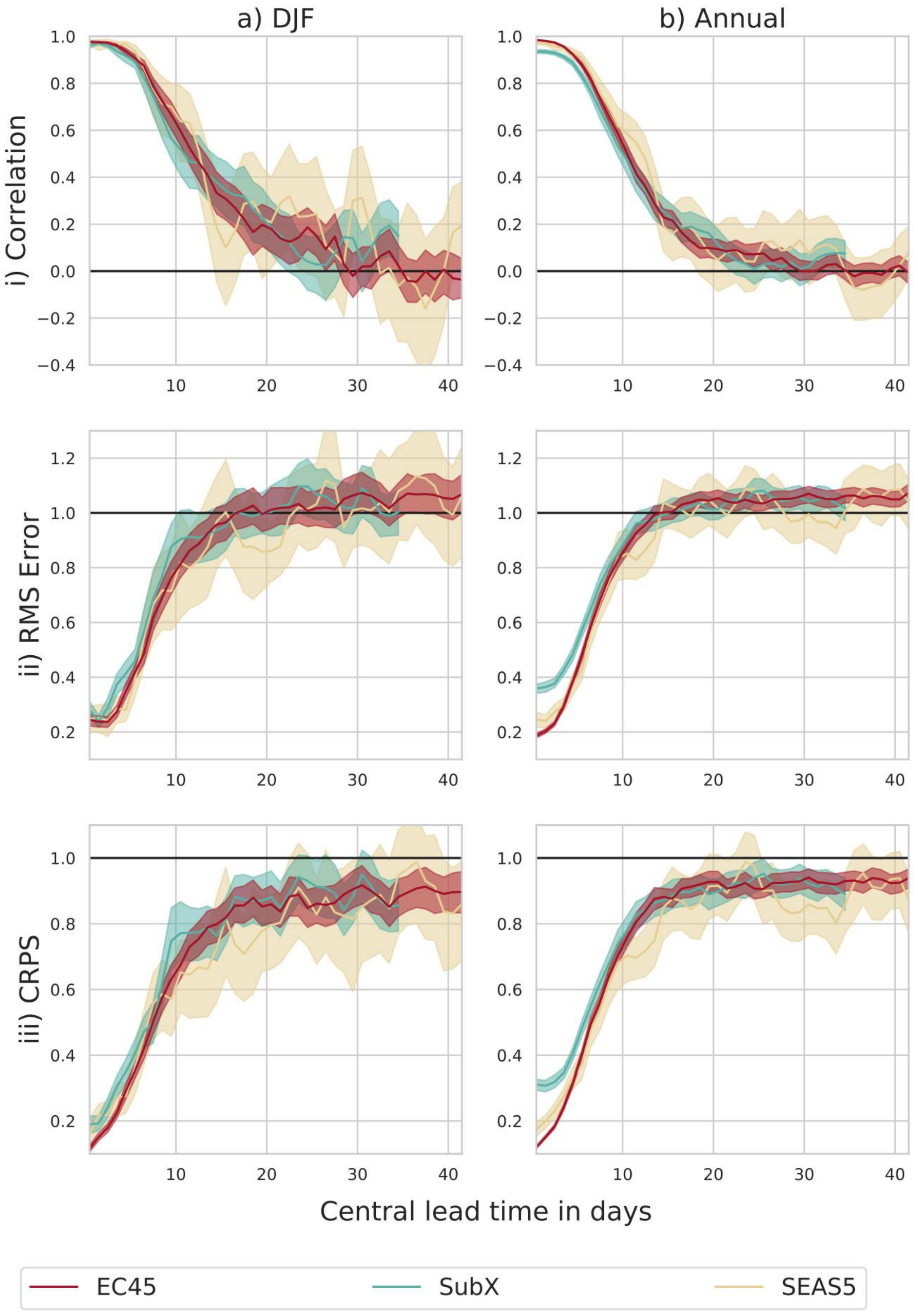}
    \caption{Measures of ensemble forecast skill for daily French 2-metre temperature anomalies during DJF \textbf{a)} and annually averaged \textbf{b)}. Correlations \textbf{i)} and RMS error \textbf{ii)} are with respect to the ensemble mean, whereas the CRPS \textbf{iii)} is a fully probabilistic score. RMSE and CRPS have been normalised by the seasonal climatological variance estimated from ERA5. Shaded regions represent the first standard deviation of bootstrap resampling over forecast initialisation dates.}
    \label{fig:skill_scores}
\end{figure}

\begin{figure}
    \centering
    \includegraphics[width=0.95\linewidth]{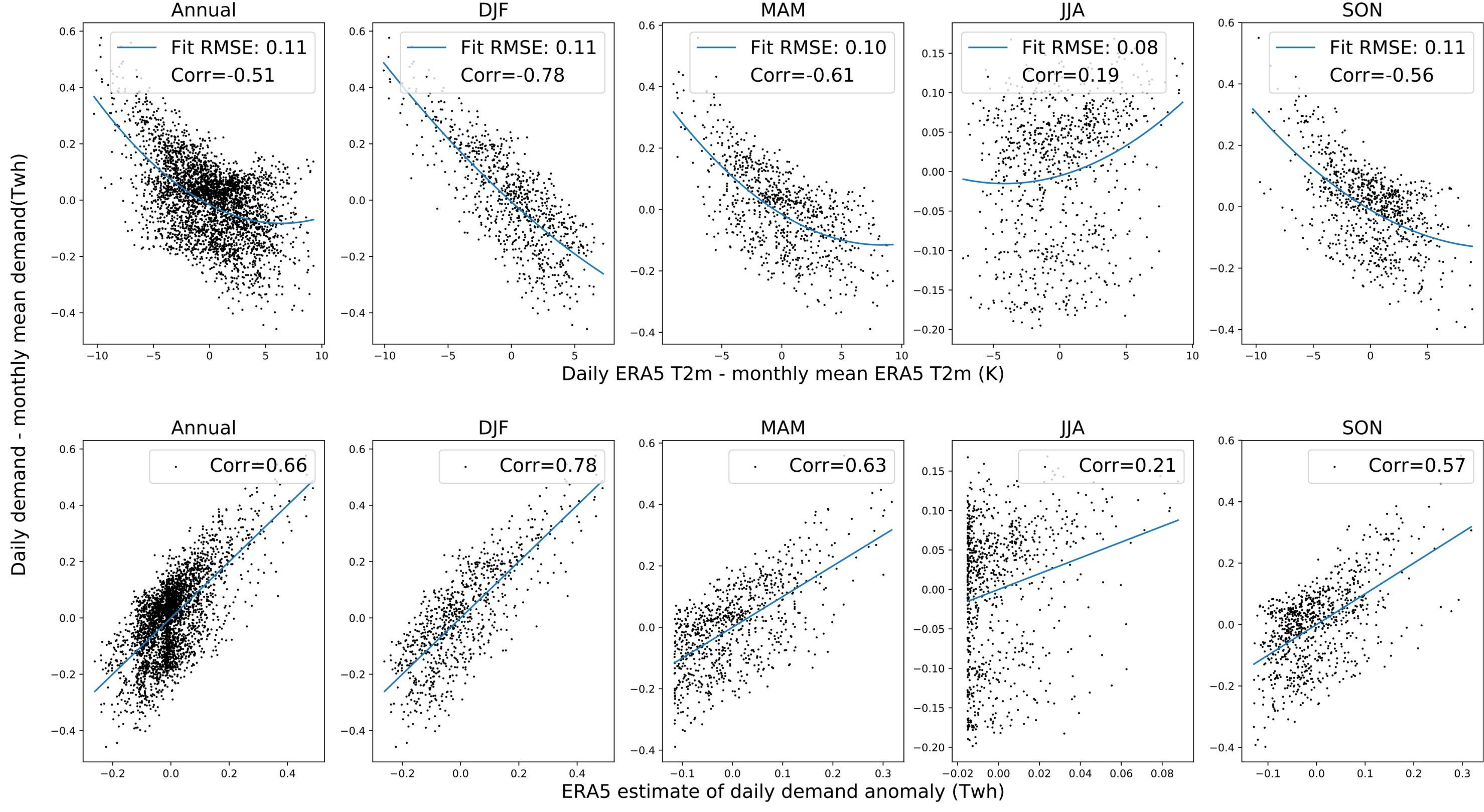}
    \caption{Top, anomalies of daily ERA5 surface temperature against anomalies of daily energy demand, with anomalies calculated with respect to the mean value for each calendar month. The solid line shows the best quadratic fit to the data. Bottom, ERA5 estimates of daily energy demand anomaly, calculated using the quadratic fit above, plotted against the same energy demand anomaly series.}
    \label{fig:fillin1}
\end{figure}
\begin{figure}
    \centering
    \includegraphics[width=0.95\linewidth]{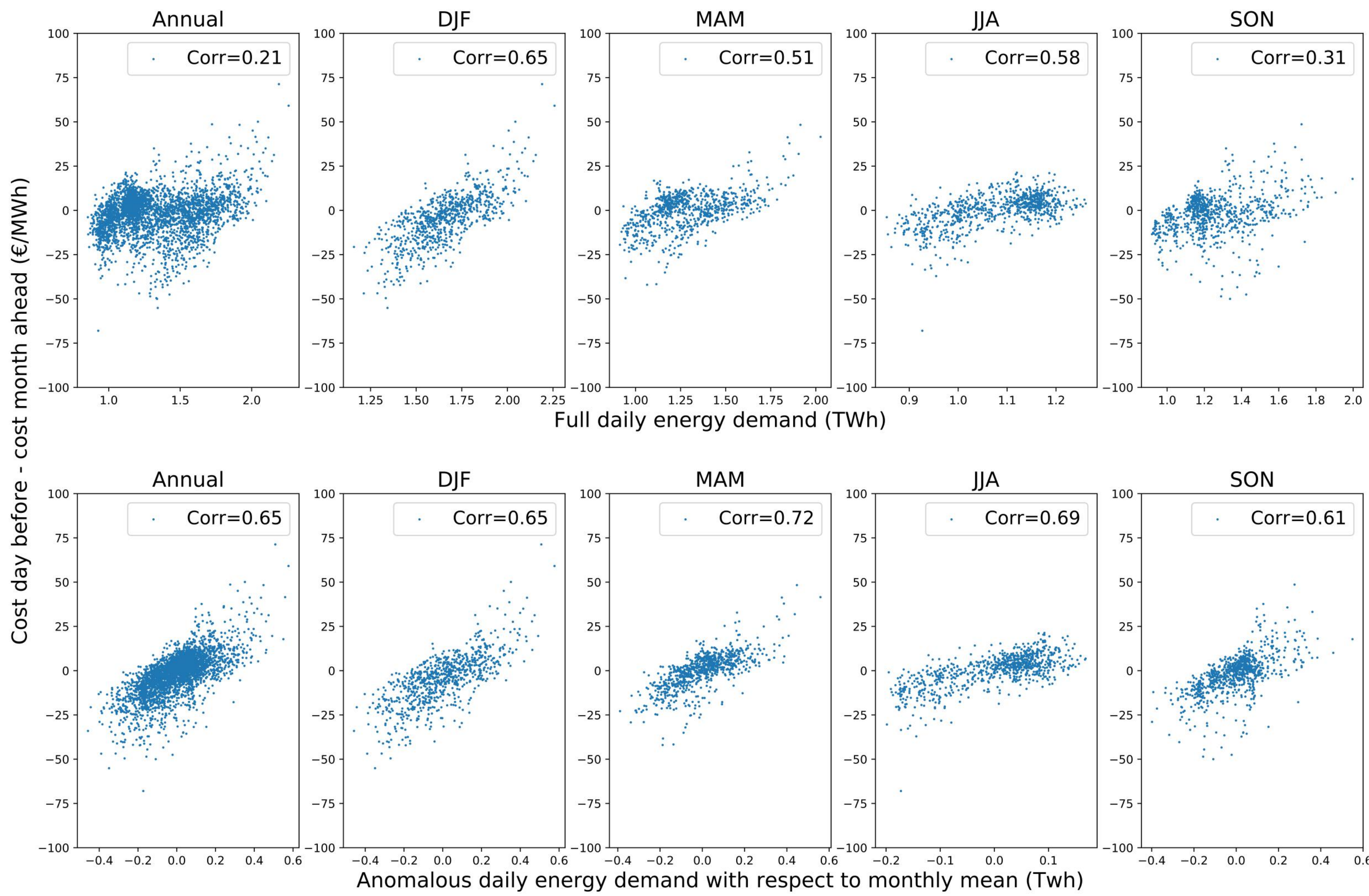}
    \caption{Top, Daily french energy demand against the difference in price between buying energy the day before delivery and buying power at the end of the previous month, termed the month-ahead cost saving. Bottom, the daily demand anomaly with respect to the mean value for each calendar month against the same month-ahead cost saving as above.}
    \label{fig:fillin2}
\end{figure}

\graphicalabstract{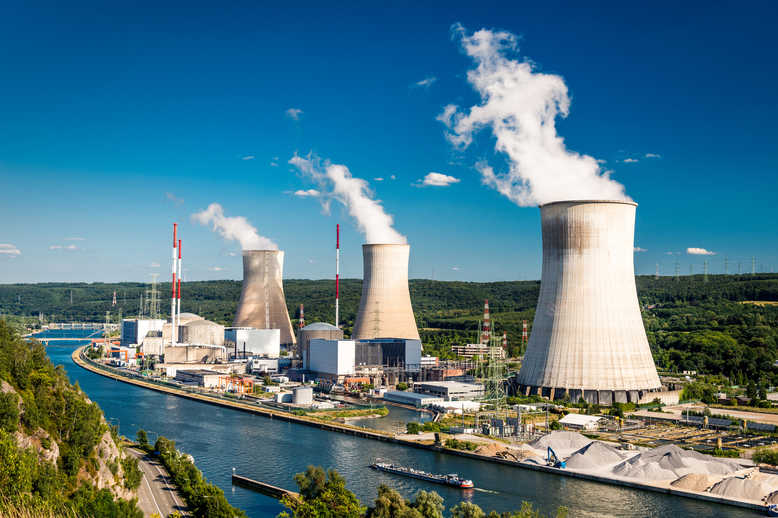}{Sub-seasonal forecasting aims to extend atmospheric predictability out 3-6 weeks into the future. We quantify the value these forecasts actually provide for a real-world example based on French energy trading. We compare the value shown by our novel real-world metric to traditional meteorological skill scores, and highlight key differences out beyond two weeks. }

\end{document}


\title{Supplementary Material for 'Beyond skill scores: exploring sub-seasonal forecast value through a case study of French month-ahead energy prediction.'}
\maketitle
\section{Stationarity of T2m}

Because two different data windows are used in this work, 1999-2018 for surface temperature and 2010-2018 for energy demand and price, it is important to compare the forecast skill of the earlier time period 1999-2009 to the later time period 2010-2018 in order to see if results using the two different data windows are comparable. We do this by computing daily correlation, RMSE, and CRPS for both the earlier and later time period at each lead time and plot the difference between them (later period skill minus the earlier period skill). To determine if this difference is significant we use a bootstrap estimate computed by randomly splitting all forecasts for each lead time into two groups, and taking the difference of one from the other. This was done 500 times for each lead time, and we consider a significant difference in skill to occur if the difference falls either beneath the 2.5th percentile or above the 97.5th percentile of the bootstrap distribution.

\begin{figure}
    \centering
    \includegraphics[width=0.95\linewidth]{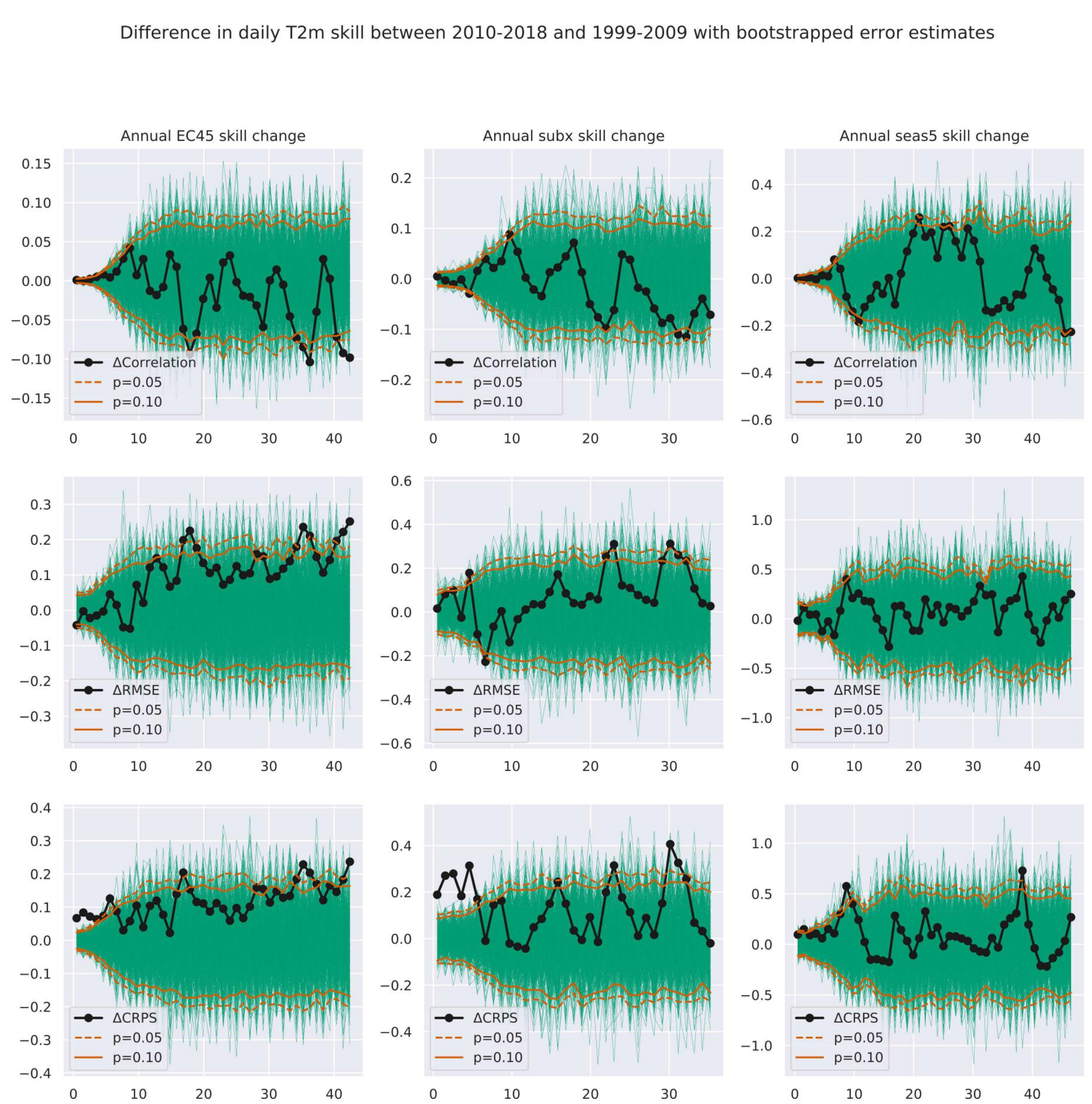}
    \caption{The difference in skill for daily forecasts over all dates of daily T2m over France between the period 2010-2018 and 1999-2009. The black line in each case shows the calculated difference, while thin green lines show the result of 500 bootstrapped differences between randomly selected parts of each forecast dataset. Thick orange lines show confidence intervals at the p=0.1 level (i.e. marking the 5th and 95th percentiles), while dashed orange lines show confidence intervals at the p=0.05 level (i.e. marking the 2.5th and 97.5th percentiles)}
    \label{fig:Annual_1d}
\end{figure}

\begin{figure}
    \centering
    \includegraphics[width=0.95\linewidth]{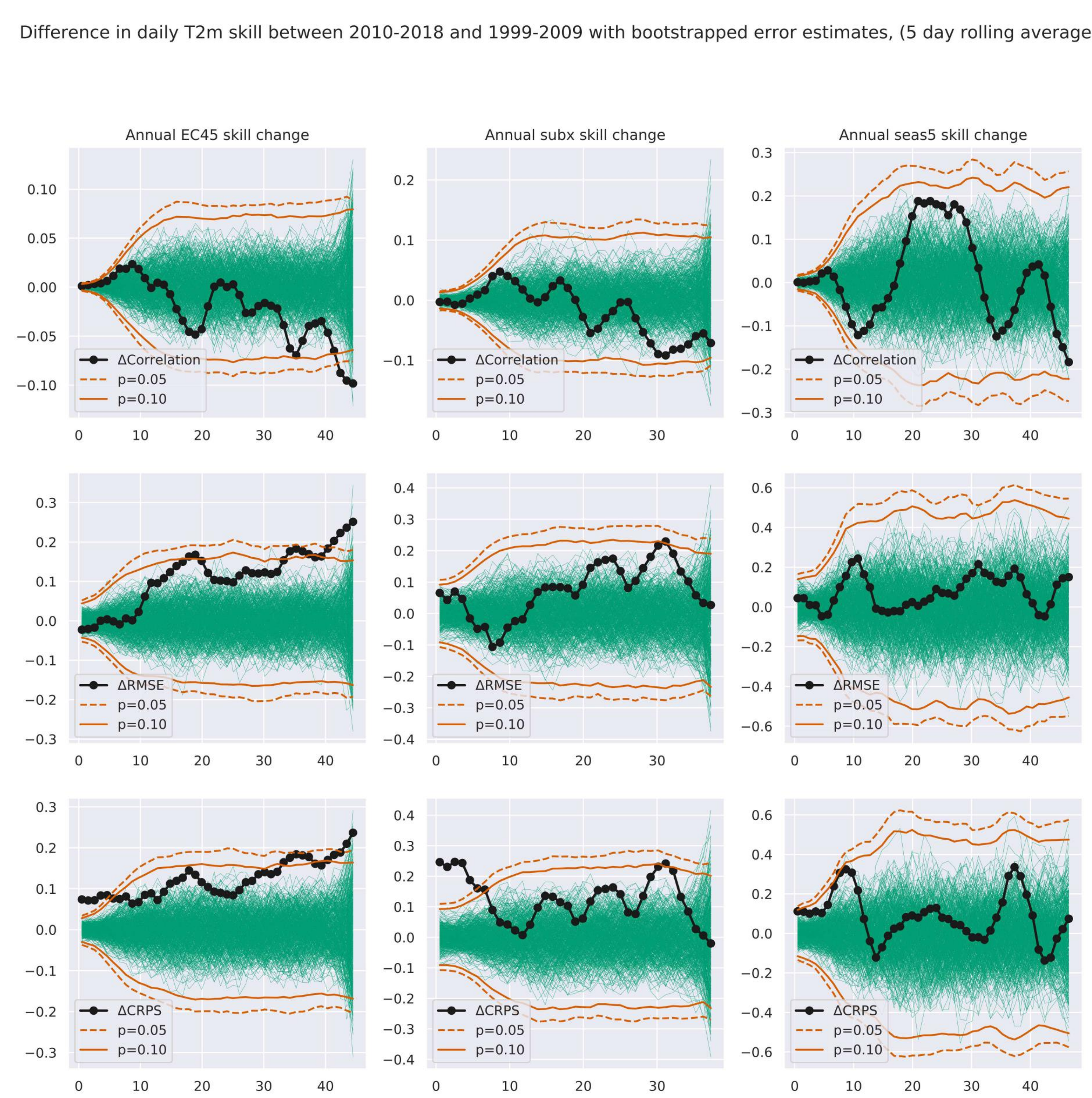}
    \caption{As in figure \ref{fig:Annual_1d}, but showing 5-day average differences. Confidence intervals are calculated from the smoothed bootstraps, not smoothed versions of the intervals in  figure \ref{fig:Annual_1d}}
    \label{fig:Annual_5d}
\end{figure}

\begin{figure}
    \centering
    \includegraphics[width=0.95\linewidth]{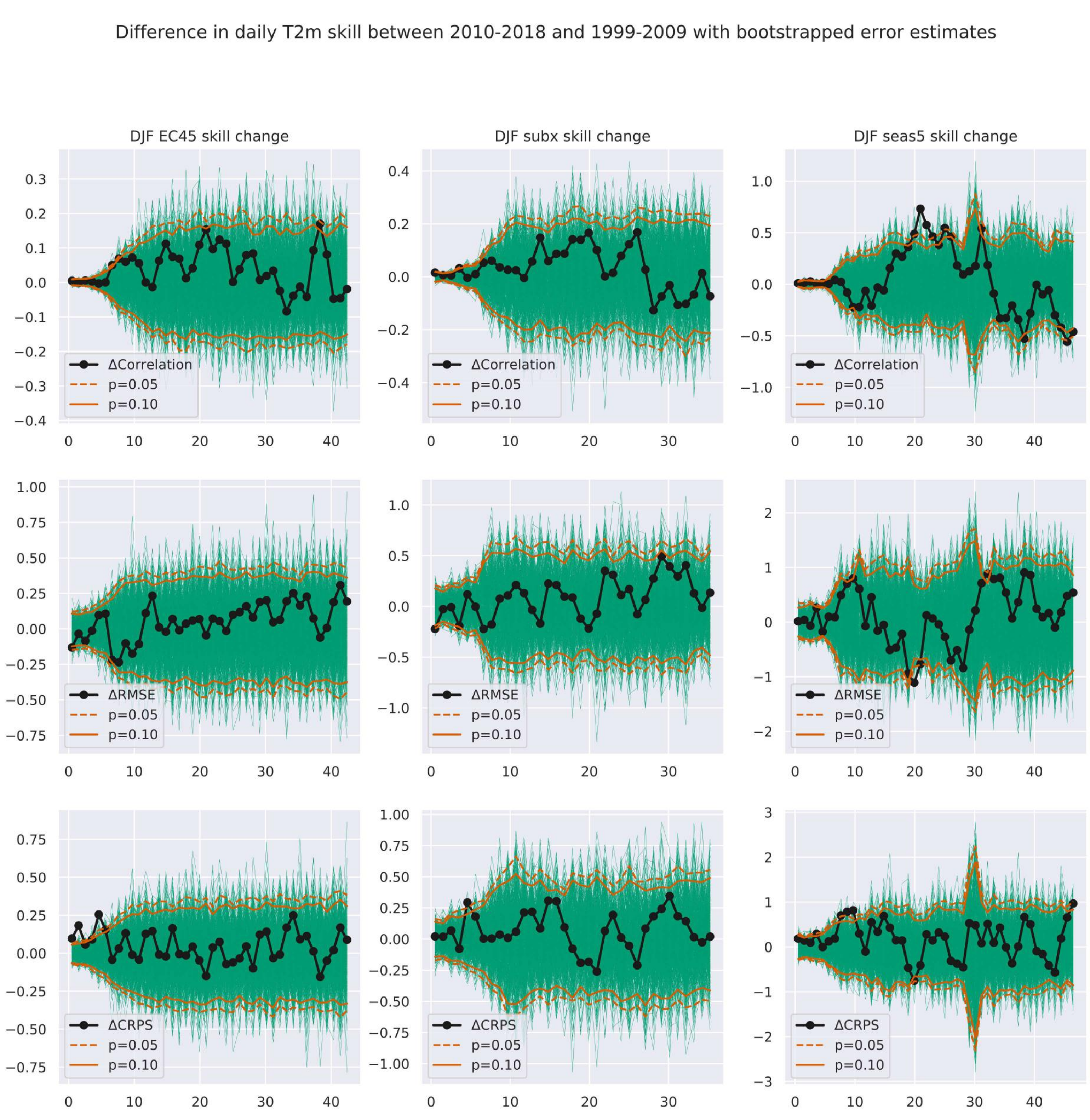}
    \caption{As in figure \ref{fig:Annual_1d} but restricted to only forecasts for DJF}
    \label{fig:DJF_1d}
\end{figure}

\begin{figure}
    \centering
    \includegraphics[width=0.95\linewidth]{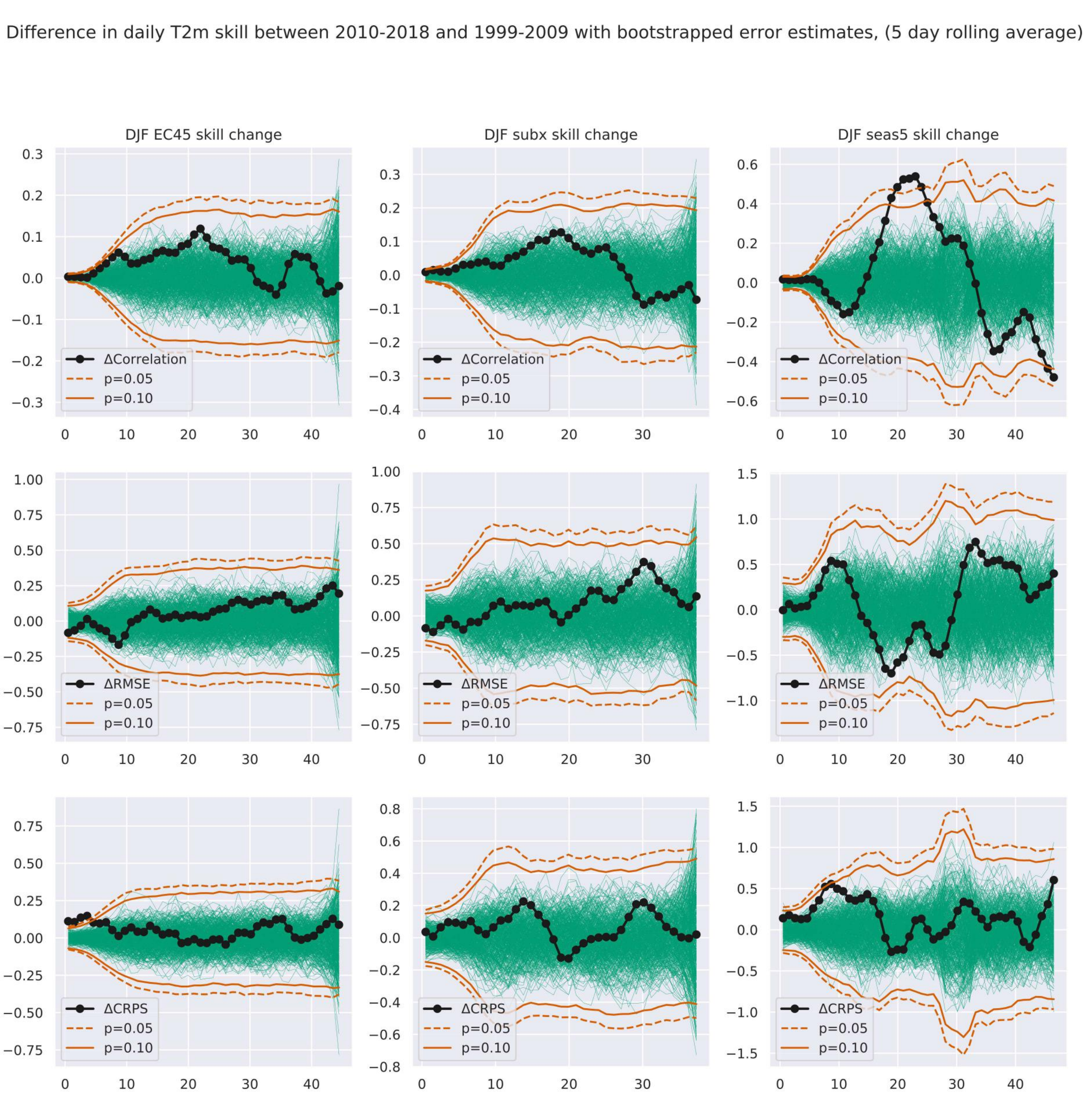}
    \caption{As in figure \ref{fig:Annual_5d} but restricted to only forecasts for DJF}
    \label{fig:DJF_5d}
\end{figure}

We show results for forecasts over the whole year in figure \ref{fig:Annual_1d}, and for DJF only in fig \ref{fig:DJF_1d}. These results are quite noisy, and as it is hard to imagine any physically meaningful reason why there might be a significant skill difference at e.g. day 20 between the late and early periods, but not on days 19 or 21, we also apply a 5-day smoothing window to these differences, as shown in figures \ref{fig:Annual_5d} and \ref{fig:DJF_5d} for the whole year and for DJF only respectively. The choice of a 5-day window is relatively arbitrary, but sensitivity tests with windows of width 3, 7 and 10 showed qualitiatively similar results. For these smoothed differences the EC45 system shows significantly worse skill annually in the later period after day 42, but at this point in the forecast daily errors are already well saturated. Both EC45 and SubX show higher CRPS on an annual basis during week 1 in the later period, indicating lower probabilistic skill, and this can also be seen in the DJF only results for EC45 at reduced magnitude. As mostly we are in the skill at weeks 2-4 in this work, and due to the relatively minor differences visible, we conclude that meaningful comparisons can be made between skill scores calculated on the two different data windows.

\section{Land Mask}

\begin{figure}
    \centering
    \includegraphics[width=0.95\linewidth]{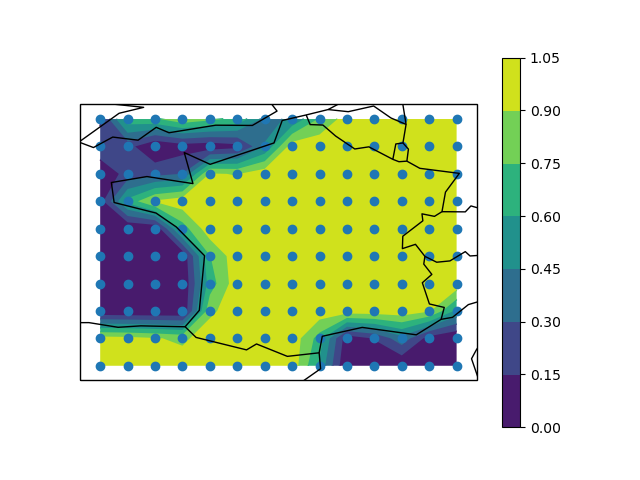}
    \caption{A map of the gridpoints included in the calculation of the box area mean used in the main text. Blue points mark the location of gridpoints at 1 degree resolution, with the contribution of each value to the area mean weighted by the value of the land sea mask at that point, shown in coloured contours. Land points are weighted 1, ocean points are weighted 0, and coastal points are weighted fractionally, as the land sea mask has been regridded from a 0.25 degree resolution map.}
    \label{fig:LM1}
\end{figure}

\begin{figure}
    \centering
    \includegraphics[width=0.95\linewidth]{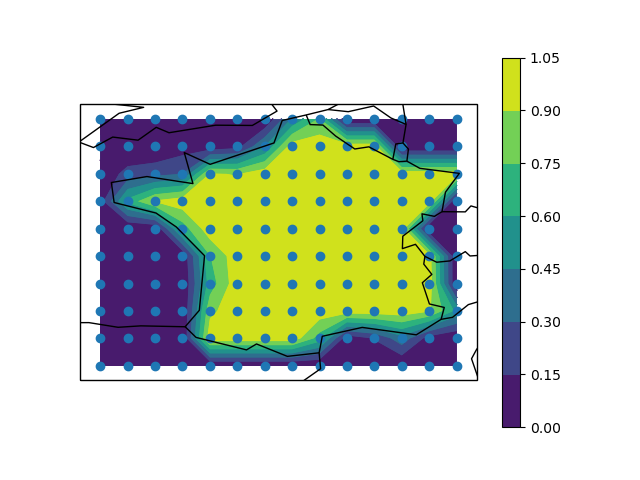}
    \caption{As in figure \ref{fig:LM1} but showing a more precise mask that removes points not within the geographical borders of France.}
    \label{fig:LM2}
\end{figure}

\begin{figure}
    \centering
    \includegraphics[width=0.95\linewidth]{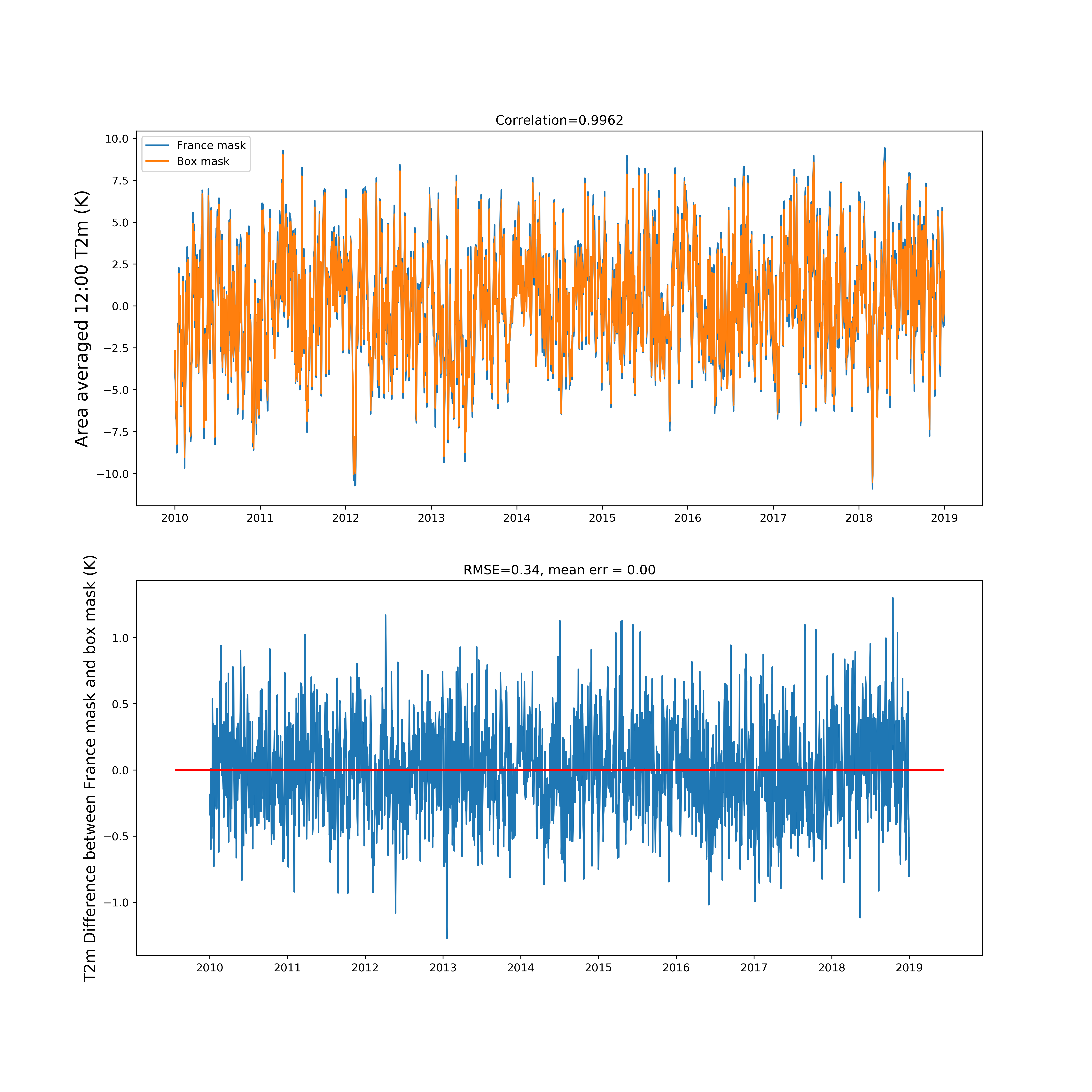}
    \caption{Top, the seasonally detrended, area averaged T2m series both for the France mask in blue (as in figure \ref{fig:LM2}) and the box mask in orange (as in figure \ref{fig:LM1}). The correlation of the two series is 0.9962. Bottom, the difference between the France mask mean and the box mask mean, with red line marking the x axis.}
    \label{fig:LM3}
\end{figure}

In our work we calculate a scalar average temperature over France by applying a land-sea mask to our gridpoint data and averaging over the box [5W-8E,42N-51N]. This was due primarily due to pre-availability of forecast time series using this box mean which had been computed for another project.  As can be seen in figure \ref{fig:LM1} this choice of region does in fact include some points from neighbouring countries. We might consider what effect this has on our results, compared to using a more precise land mask such as that in \ref{fig:LM2}. We compute area means of ERA5 reanalysis T2m over the period 2010-2019 for the box mask of figure \ref{fig:LM1} and the France mask of figure \ref{fig:LM2}, deseasonalise and make a comparison between them in figure \ref{fig:LM3}. The correlation between the two time series is 0.9962, meaning the cruder box mean series is capturing essentially all the variation present in the French temperature series. The RMS error of 0.34 is approximately 10\% of the standard deviation of each of the detrended anomaly series. We therefore feel confident that we lose nothing by using the box mean approach over a more precise France mean.

